# Discrete Gompertz and Generalized Logistic models for early monitoring of the COVID-19 pandemic in Cuba

**Abstract**

The COVID-19 pandemic has motivated a resurgence in the use of phenomenological growth models for predicting the early dynamics of infectious diseases. These models assume that time is a continuous variable, whereas in the present contribution the discrete versions of Gompertz and Generalized Logistic models are used for early monitoring and short-term forecasting of the spread of an epidemic in a region. The time-continuous models are represented mathematically by first-order differential equations, while their discrete versions are represented by first-order difference equations that involve parameters that should be estimated prior to forecasting. The methodology for estimating such parameters is described in detail. Real data of COVID-19 infection in Cuba is used to illustrate this methodology. The proposed methodology was implemented for the first thirty-five days and was used to predict accurately the data reported for the following twenty days.

**Acronyms and Symbols**

$G$: Gompertz.
$GL$: Generalized Logistic.
$C(t)$: Cumulative number of infections at continuous time $t$.
$\{C_n^0\}_{n=1}^{N_0}$: Discrete time series of the cumulative number of infections.
$\{C_n\}_{n=1}^{N}$: Discrete time series of the cumulative number of infections after averaging with a 7 day long moving window.
$\Pi$: Model-specific parameter vector.
$K$: Carrying capacity.
γ: Intrinsic growth rate.
μ: Deceleration of growth.

## 1. Introduction

Growth models are characterized by first-order ordinary differential equations that depend on certain parameters that can be estimated using real data through optimization methods. The interrelations and generalizations of the basic growth models (Malthus, Verhulst logistic, Richards and Gompertz) can be found in (Tsoularis & Wallace 2002, p. 21). A historical development of these models may be found, for instance, in (Kingsland 1982, p. 29). In (Bürger et al. 2021, p. 108558-4) a general methodology based on phenomenological growth models is presented for the application of statistics to various problems in medicine and biology.

Growth models have been used to describe the early stage of several epidemics, as can be seen in (Brauer et al. 2019, p. 113). In (Zhou & Yan 2003, p. 1608) the simple Richards model (R) is used for describing the dynamics of cumulative cases by Severe Acute Respiratory Syndrome in Singapore, Hong Kong and Beijing in 2003. In (Pell et al. 2018, p. 62) the logistic growth model (L) and the generalized Richards model (GR) are calibrated with data of 2015 Ebola epidemic in West Africa. In (Shanafelt et al. 2018, p. 338) a generalized-growth model is used to investigate the 2001 Foot-and-Mouth Disease epidemic in the UK. In (Dinh et al. 2016, p. 20) the L model is applied for analyzing the dynamics of Zika outbreak in Florida, USA, 2016. In (Zhao et al. 2019, p. 2) the L, the Gompertz (G) and the GR models are used for the early prediction of the outbreaks of the Zika epidemic in various states of Brazil, during 2015-2016.

Recently, several scientists have used growth models to study the spread of the coronavirus disease (COVID-19) that has plagued humanity since December 2019. In (Malavika et al. 2021, p. 26) the L model is used for short term forecasting the COVID-19 epidemic in India. In (Shen2020, p. 582) the L model is used for studying the propagation of COVID-19 in China, South Korea and Iran. In (Wu et al. 2020, p. 1561) the L model, the generalized logistic (GL), the GR and the generalized growth models are calibrated with the reported number of COVID-19 infected cases for the whole of China, 29 Chinese provinces, and 19 countries. In (Vasconcelos et al.2020, p. 100121) the GR model is applied to the COVID-19 fatality curves from several countries. In (Jain et al. 2020, p. 784) the L, the GL and the generalized growth models are applied to identify the initial climbing growth period of COVID-19 outbreak in India from Apr 10, 2020, to Apr 20, 2020. In (Roosa et al. 2020ª, p. 596) the L and the GR models are used to generate short-term forecasts of cumulative reported cases in Guangdong and Zhejiang, China. In (Roosa et al. 2020b, p. 256), the GL and GR models are validated during earlier outbreaks to generate and evaluate short-term forecasts of the cumulative number of confirmed reported cases in Hubei province, China. In (Wu et al. 2021, p. 603001) the generalized growth models are applied to quantify the evolution of COVID-19 in different countries. Simple phenomenological models (generalized growth model and logistic) are used in (López et al. 2021, p. e0253004) to characterize the two first outbreak waves of COVID-19 in Spain. In (Pincheira& Betancor 2021, p. 100486) it is shown that a semi-unrestricted version of the generalized growth model outperforms the traditional in several countries when predicting the number of infected people at short period.

The mathematical models used in the aforementioned works are represented by first-order differential equations on a continuous time domain:

$$C'(t) = f(C, t, \boldsymbol{\Pi}), \quad t \geq 0, \qquad (1)$$

where $C$ is the cumulative number of infected cases at time $t$ and $\boldsymbol{\Pi}$ is a vector of parameters that is specific for each model. $f$ is a known real function that depends on $C, t$ and the parameters of the vector $\boldsymbol{\Pi}$. For instance, for the Gompertz model, $f(C, t, \boldsymbol{\Pi}) = \gamma C \ln(K/C)$ as can be seen in Table 1. $C(t)$, on the other hand, is usually a curve with exponential or subexponential behavior at low values and with an inflexion point between its minimum value and the upper asymptote (Tsoularis & Wallace 2002, p. 21).

In this work, we are interested in the study of first-order *difference* equations that can be derived from Eq.(1), that is:

$$C_{n+1} = C_n + hf(C_n, n, \boldsymbol{\Pi}), \quad n = 1, 2, \ldots, \quad (2)$$

where $C_n$ represents the cumulative number of infected cases at the $n$-th measurement of the (discrete) data and $h$ represents the time difference between consecutive readings (assuming it is constant). One can imagine that, since the number of infected people is typically reported at regular time intervals (daily, weekly, et cetera) the discrete models are suitable for describing the dynamics of the disease. In (May 1976, p. 459), an iconic review on the theoretical and applied scope of these equations is reported. The objective of this work is to illustrate the potential of the discrete phenomenological models associated to continuous time first-order differential equations, as an alternative for estimating the early spread of epidemics. A methodology for a statistical study of the related parameters is described, the application of which is exemplified based on initial data from the COVID-19 in Cuba.

The work is structured as follows: In section 3 the methodology is described in detail. In section 4 the methodology is applied for the case of Gompertz and Generalized Logistic models. Section 5 is devoted to some concluding remarks.

## 2. Models and Methodology

*Models*

The standard way of obtaining the difference equation (2) from the continuous one (1) is through the transformation:

$$C(t) \to C_n, \quad C'(t) \to (C_{n+1} - C_n)/h, \qquad (3)$$

Where the continuous function $C(t)$ is replaced by the time series $\{C_n\}_{n=1}^{N_0}$, and $h$ is the time step between recorded values (we are assuming that it is constant). The model is specified by the choice of the function $f(C, t, \Pi)$ where the vector of parameters $\Pi$ may have different lengths (Table 1).

**Table 1.** *G and GL models, their continuous and discrete equations and their vectors of parameters.*

|  | $C'(t) =$ | $C_{n+1} =$ | $\Pi =$ |
|---|---|---|---|
| (G) | $\gamma C \ln(K/C)$ | $C_n + h\gamma C_n \ln(K/C_n)$ | $(K, \gamma)$ |
| (GL) | $\gamma C^\mu (1 - C/K)$ | $C_n + h\gamma C_n^\mu (1 - C_n/K)$ | $(K, \gamma, \mu)$ |

The parameter $K$ is called the carrying capacity, it corresponds to the upper asymptote of the growth curve. The parameter $\gamma$ is the intrinsic growth rate and the parameter $\mu$ is called deceleration of growth when $0 < \mu < 1$ (Viboud et al. 2016, p. 28). These two latter parameters characterize the growth at low values, which is exponential (if $\mu = 1$) or subexponential (if $0 < \mu < 1$).

*Methodology*

A two-stage procedure was developed. In the first part:

1. The time series $\{C_n^0\}_{n=1}^{N_0}$ of daily reported cumulative number of cases is averaged with a 7 day long moving window, resulting in the time series $\{C_n\}_{n=1}^{N}$, where $N = N_0 - 6$. This smoothing has the effect of eliminating factors like delays in reporting due to the accumulation of unreported cases. This kind of window averaging has been used before in COVID-19 literature (León-Mecías & Mesejo-Chiong 2020, p. 23).

2. From the time series $\{C_n\}_{n=1}^{N}$ the first $N_1$ elements are taken to be used to train the model (calibration period) and form the list of pairs $\{(C_n, C_{n+1})\}_{n=1}^{N_1-1}$.

3. The list of pairs is fitted to the model $Y = X + hf(X, n, \Pi)$, that corresponds to the equation (2), by using the function NonlinearModelFit from the software Wolfram Mathematica. From here we obtain the optimal vector $\Pi^*$ of the parameter vector $\Pi$. The constant value $h = 1$ was chosen.

4. The predicted time series $\{C_n\}_{n=1}^{N}$ is computed through the RecurrenceTable function using the optimal vector $\Pi^*$ and $C_1^* = C_1$.

In the second part of the methodology, the quality of the fit is evaluated by estimating the error through bootstrap (see, for instance, Efron & Tibshirani 1994, p. 45) on $M$ realizations ($M = 10^5$ in this work). Each realization corresponds to one time series $\{C_{nm}\}_{n=1}^{N}$ with $n = 1, 2, \ldots, M$. The following steps illustrate how to generate one realization:

5. For each one of the $n = 1, 2, ..., N$ an error $e_{nm}$ is simulated. This is equal to a Poisson random number with mean $\lambda_n = C^*_{n+1} - C^*_n$.

6. The parameter vector $\boldsymbol{\Pi}^{**}_m$ is estimated as in step 3. above from the time series $\{C^*_n + e_{nm}\}$.

7. For each model, the estimation of the error bounds, mean values and standard deviations is done over the set of parameter vectors $\{\boldsymbol{\Pi}^{**}_m\}_{m=1}^M$.

**3. Results**

In this section the results for the case of the Gompertz and Generalized Logistic models are shown. The data corresponds to the cumulative number of cases of COVID-19 during the first 61 days of the pandemic in Cuba (from March 11, 2020 to May 10, 2020) and was taken from the official government site https://covid19cubadata.github.io/. The list of the initial data used in the calculations is {3, 3, 4, 4, 4, 5, 7, 11, 16, 23, 33, 38, 46, 55, 65, 78, 117, 137, 168, 184, 210, 231, 267, 286, 318, 348, 394, 455, 513, 562, 618, 667, 724, 764, 812, 860, 921, 984, 1033, 1085, 1135, 1187, 1233, 1283, 1335, 1367, 1387, 1435, 1465, 1499, 1535, 1609, 1647, 1666, 1683, 1701, 1727, 1739, 1752, 1764, 1781}. The window-averaged data consists of a time series of 55 days.

The optimal values $\boldsymbol{\Pi}^*$ of the parameter vector after step 3 of the methodology are reported in Table 2.

**Table 2.** *Optimal values of the parameter vector $\boldsymbol{\Pi}^*$.*

|      | $K^*$   | $\gamma^*$ | $\mu^*$  |
|------|---------|------------|----------|
| (G)  | 2446.64 | 0.0594009  | -        |
| (GL) | 1825.74 | 0.588639   | 0.759485 |

The mean values and standard deviations of the set of parameter vectors after step 7 of the methodology are reported in Table 3.

**Table 3.** *Mean values and standard deviations $\boldsymbol{\sigma}$ of the parameter vector $\boldsymbol{\Pi}^*$ after bootstrap.*

|       | $\bar{K}$ | $\sigma_K$ | $\bar{\gamma}$ | $\sigma_\gamma$ | $\bar{\mu}$ | $\sigma_\mu$ |
|-------|-----------|------------|----------------|-----------------|-------------|--------------|
| (G)   | 2425.44   | 166.877    | 0.0602047      | 0.00253007      | -           | -            |
| (GL)  | 1836.31   | 214.559    | 0.5925         | 0.10807         | 0.762901    | 0.0378191    |

In Figure 1 we show the window-averaged data in dots vs. the forecasted line obtained by using NonlinearModelFit with the Gompertz model. The calibration period was $N_1 = 35$ days on the 55 first days of the early onset of the pandemic in Cuba. The upper and lower bounds for the 95% confidence interval for each parameter were estimated via bootstrap as described above. These lower and upper bounds allow us to have a *fan* of possible trajectories of $C_n$ for the forecasted period. The histograms for each one of the parameters are shown in Figure 2. The same graphics using the generalized logistic model are shown in Figure 3. The normality assumption is appropriate, as we can note, from the central peak and symmetry of the histograms. The implemented code for Gompertz model can be found in https://notebookarchive.org/2022-05-1fu9m2t.

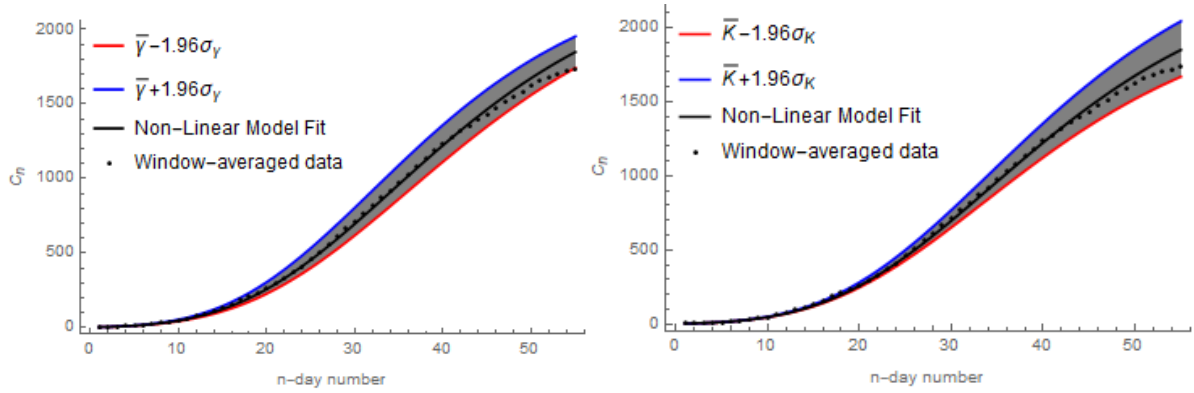

**Figure** 1. Real data (window-averaged), forecasted time series with NonlinearModelFit and upper and lower bounds of 95% confidence interval for both parameters of the Gompertz model.

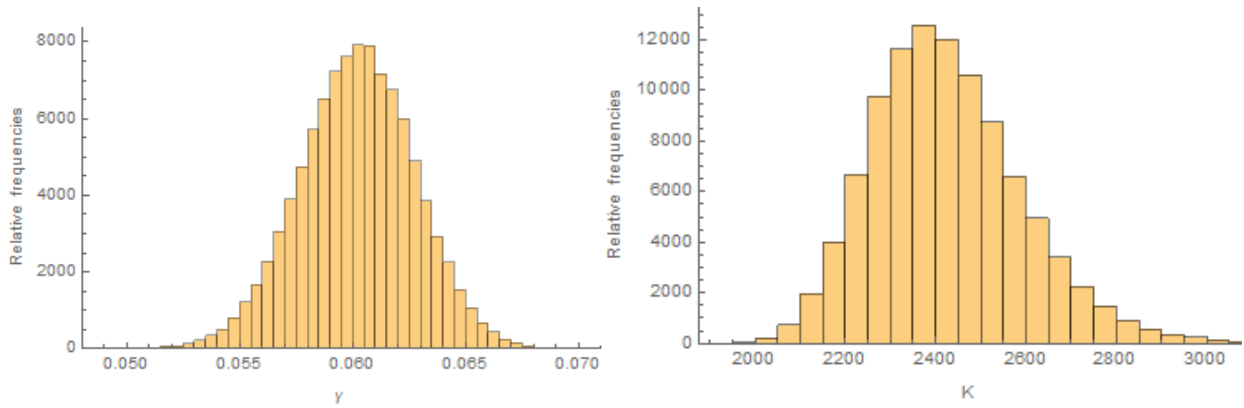

**Figure** 2. Histograms of the parameters of the Gompertz model generated via Bootstrap

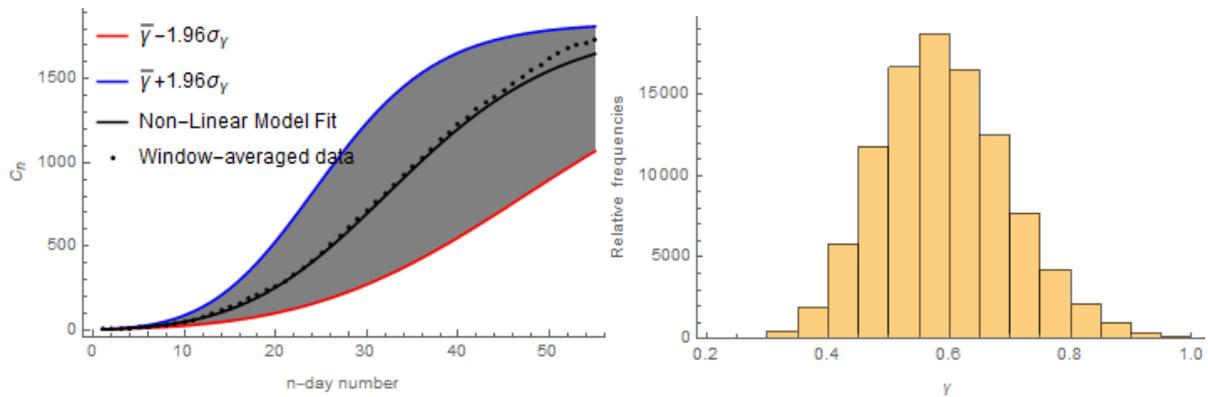

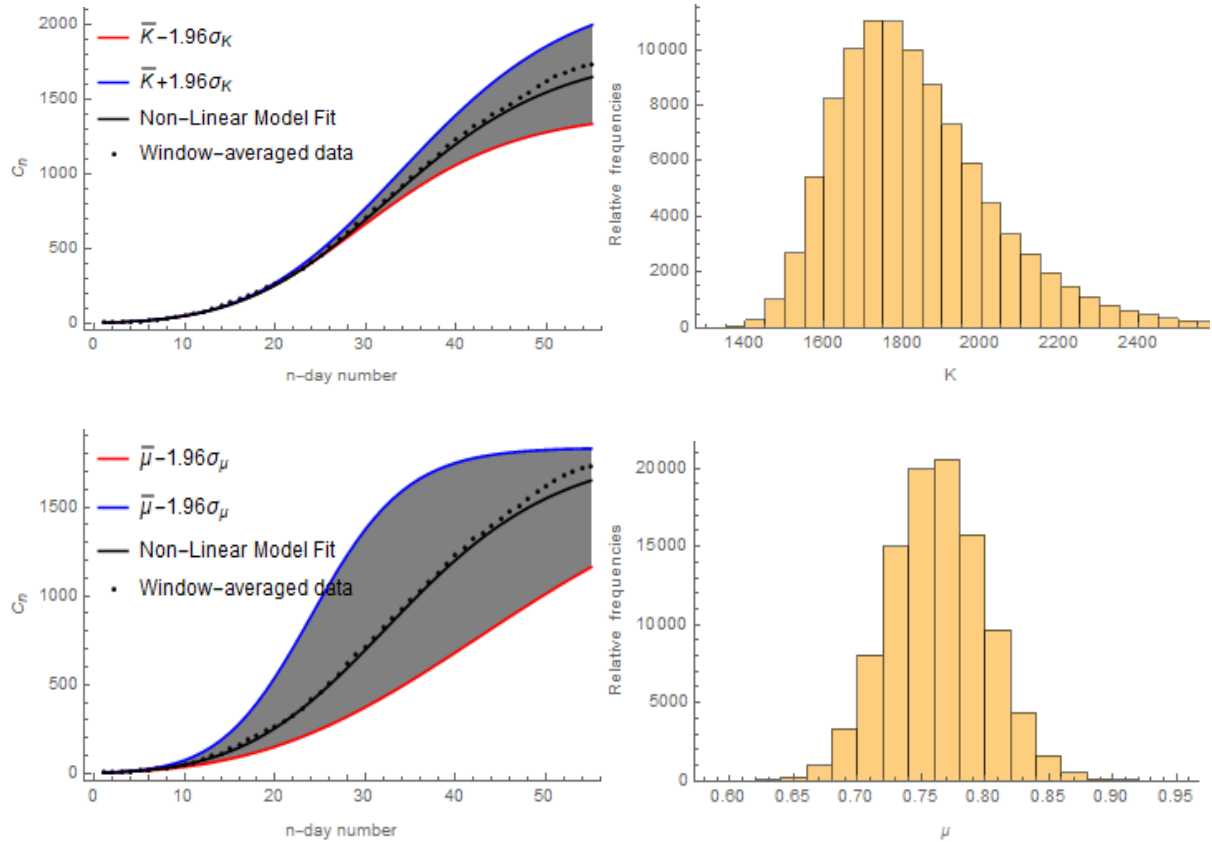

**Figure** 3. Real data (window-averaged), forecasted time series with NonlinearModelFit and upper and lower bounds of 95% confidence interval for both parameters of the Generalized Logistic model (left). Histograms of the parameters of the Generalized Logistic model generated via Bootstrap (right).

## 4. Conclusions

A methodology based on discrete phenomenological models was illustrated in detail as an alternative to estimate the early evolution of epidemics. The methodology was applied to the Gompertz and Generalized Logistic models which depend on two and three parameters, respectively. The data used in the computations were the initial real data (since March 11, 2020 to May 10, 2020) on the evolution of COVID-19 in Cuba.

The numerical implementation was carried out in Wolfram Mathematica in a simple and direct way. The procedure consists of two fundamental stages. In the first, the optimized parameters are obtained, while in the second, a statistical analysis is performed to evaluate the error of the optimization process. The proposed methodology was implemented for the first thirty-five days, being able to predict with very good precision the data reported for the following twenty days. The results show that the proposed methodology could be useful to guide decision-making that facilitates mitigating the early effects of the epidemic.

## 5. Acknowledgements

ROC would like to thank financing from FENOMEC, UNAM. Financial support from the PAPIIT DGAPA UNAM IN10182 project is gratefully acknowledged. Authors would also like to thank for the useful suggestions received from the referees.

**Authors contributions**

All authors contributed equally to this work.

**Conflict of interest**

All authors declare no conflicts of interest in this paper.